\documentclass[fleqn,twoside]{article}
\usepackage{espcrc2,cite,graphicx,amsmath,amsfonts,amssymb,mathrsfs,feynmp}
\usepackage[figuresright]{rotating}

\newcommand{\ie}{{\it i.e.}}

\newcommand{\eg}{{\it e.g.}}

\newcommand{\eq}{Eq.}
\newcommand{\eqs}{Eqs.}

\newcommand{\figs}{Figs.}
\newcommand{\Ref}{Ref.}

\newcommand{\Sec}{Sec.}
\newcommand{\App}{App.}

\begin{document}

\title{
\vspace*{-3cm}
\begin{flushright}
{\small TUM-HEP-457/02}
\end{flushright}
\vspace*{0.5cm}
{\bf Bilarge leptonic mixing from Abelian horizontal symmetries}}

\author{{\large Tommy Ohlsson}\address[TUM]{{\it Institut f{\"u}r Theoretische
Physik, Physik-Department, Technische Universit{\"a}t M{\"u}nchen,
James-Franck-Stra{\ss}e, 85748 Garching bei M{\"u}nchen,
Germany}}\thanks{E-mail: {\tt tohlsson@ph.tum.de}},
{\large Gerhart Seidl}\addressmark[TUM]\thanks{E-mail: {\tt
gseidl@ph.tum.de}}}
 
\begin{abstract}
\noindent {\bf Abstract} 
\vspace{2.5mm}

We construct and present a model for leptonic mixing based on
higher-dimensional operators, using the Froggatt--Nielsen mechanism,
and Abelian horizontal symmetries (flavor symmetries) of continuous
and discrete type. Our model naturally yields bilarge leptonic
mixing, coming from both the charged leptons and the neutrinos,
and an inverted neutrino mass hierarchy spectrum. The obtained values
of the parameters, \ie, the leptonic mixing parameters and the
neutrino mass squared differences, are all consistent with the
atmospheric neutrino data and the Mikheyev--Smirnov--Wolfenstein large
mixing angle solution for the solar neutrino problem.

\vspace*{0.2cm}
\noindent {\it PACS:} 14.60.Pq, 11.30.Hv\\
\noindent {\it Key words:} Leptonic mixing; Neutrino masses; Discrete
symmetries; Higher-dimensional operators
\end{abstract}

\maketitle

\section{Introduction}

Predicting the pattern of fermion masses and mixings from a
fundamental gauge theory is one of the major challenges in particle physics. In
such an approach, the observed hierarchy of the fermion masses is usually
understood in terms of some symmetry breaking interaction.
In fact, due to their plausible Majorana nature, the extreme
smallness of the neutrino masses could be associated with a violation of
the $B-L$ symmetry. Thus, the neutrinos can shed light on the origin
of the fermion masses and mixings, since most grand unified theories
(GUTs) based on SO(10) or $E_6$ and also string theories indeed
expect the $B-L$ symmetry to be broken \cite{witt00}. In building models,
it is therefore particularly important to naturally reproduce
the neutrino mass squared differences and the leptonic mixing parameters
that have been determined by the atmospheric
\cite{Fukuda:1998mi,Toshito:2001dk} and
solar \cite{Fukuda:2001nj} neutrino data.

The neutrino mass squared differences are generally defined as
$$
\Delta m_{ab}^2\equiv m_a^2-m_b^2,
$$
where $m_a$ is the mass of the $a$th neutrino mass eigenstate. We will here
assume that there are three neutrino flavors, and
therefore, three neutrino flavor states $\nu_\alpha$ ($\alpha =
e,\mu,\tau$) and also three neutrino mass eigenstates $\nu_a$ ($a =
1,2,3$). The unitary leptonic mixing matrix\footnote[1]{The leptonic
mixing matrix is sometimes called the Maki-Nakagawa-Sakata (MNS) mixing
matrix \cite{maki62}.} is then given by
$$
U = (U_{\alpha a}) \equiv {U^\ell}^\dagger U^\nu,
$$
where the unitary mixing matrix $U^\ell$ ($U^\nu$) rotates the
left-handed charged lepton fields (the neutrino fields) so that the charged
lepton mass matrix (the neutrino mass matrix) becomes diagonal. Thus,
the leptonic mixing matrix acquires contributions from both the charged
leptons and the neutrinos. These contributions usually add in a non-trivial
way (see \App~\ref{app:U}). For three neutrino flavors, in the
so-called standard parameterization, the leptonic mixing matrix reads
$$
U = {\tiny
\left( \begin{matrix} C_2 C_3 & S_3 C_2 & S_2 {\rm e}^{- {\rm i} \delta}\\
- S_3 C_1 - S_1 S_2 C_3 {\rm e}^{{\rm i} \delta} & C_1 C_3 - S_1 S_2 S_3
{\rm e}^{{\rm i} \delta} & S_1 C_2\\
S_1 S_3 - S_2 C_1 C_3 {\rm e}^{{\rm i} \delta} & - S_1 C_3 - S_2 S_3 C_1
{\rm e}^{{\rm i}\delta} & C_1 C_2 \end{matrix}\right),
}
$$
where $S_a \equiv \sin \theta_a$, $C_a \equiv \cos \theta_a$ (for $a =
1,2,3$), and $\delta$ is the physical ${\cal CP}$ phase. Here $\theta_1 \equiv
\theta_{23}$, $\theta_2 \equiv \theta_{13}$, and $\theta_3 \equiv
\theta_{12}$ are the leptonic mixing angles.
Recent results suggest that among the possible solutions to the
solar neutrino problem, the Mikheyev--Smirnov--Wolfenstein (MSW)
\cite{mikh85} large mixing angle (LMA) solution is
somewhat preferred to the MSW small mixing angle (SMA) solution, the
MSW low mass (LOW) solution, and the vacuum oscillation (VAC) solution
\cite{Fogli:2001vr,bahc01,Bahcall:2001cb}.
Actually, a global solar two flavor neutrino oscillation analysis
including the latest SNO data strongly favors the MSW LMA solution
\cite{ahma02}.
However, the MSW LMA solution excludes maximal solar mixing at the
95\% confidence level \cite{bahc01} (and now even at the 99.73\%
confidence level \cite{Bahcall:2001cb}), and therefore, it also disfavors the
scenario of so-called bimaximal mixing \cite{barg98}.
Thus, we will instead, most probably, have a {\em bilarge} mixing
scenario in which the solar mixing angle $\theta_{12}$ is large, but not
maximal, and the atmospheric mixing angle $\theta_{23}$ is
approximately maximal.
It is interesting to observe that there are very strong indications
that the leptonic mixing is large, whereas, on the other hand, it
has turned out experimentally that the quark mixing is small \cite{groo00}.

In this paper, we will investigate a model, which yields in a
technically natural way bilarge leptonic mixing, reproduces the
observed mass hierarchy of charged leptons, and leads
to an inverted neutrino mass hierarchy spectrum. This will be achieved
by generating lepton mass matrix textures, where the mixing of the
charged leptons is comparable with the mixing of the quarks and the mixing of
the neutrinos is essentially bimaximal. The striking difference
between the bilarge leptonic mixing and the small quark mixing will
then be accounted for by the neutrinos (mainly) and the charged
leptons (partly).

This paper is organized as follows: In \Sec~\ref{sec:chargedleptons},
we will introduce a model by adding to the standard model (SM) a set of extra
fields and horizontal symmetries that will give rise to specific effective
Yukawa interactions for the charged leptons. Next, minimizing the
corresponding scalar potential, we will naturally obtain a mass matrix
texture for the charged leptons, which is in agreement with
experimental data.
In \Sec~\ref{sec:neutrinomassmatrix}, we will extend the representation content
of the model in order to also obtain a realistic mass matrix texture
for the neutrinos.
(In \App~\ref{app:U}, we will explicitly calculate the leptonic mixing
angles coming from the diagonalizations of the charged lepton and
neutrino mass matrices, respectively.)
Finally, in \Sec~\ref{sec:S&C}, we will present a summary as well as our
conclusions.

\section{Charged leptons}
\label{sec:chargedleptons}

\subsection{Horizontal symmetries}
\label{sec:h_symm}

We will here consider an extension of the SM in which
the lepton masses arise from higher-dimensional operators
\cite{wein79} via the Froggatt--Nielsen mechanism
\cite{frog79}. (For recent studies, see, \eg,
\Ref~\cite{Babu:2001ex}.) We will write, in a self-explanatory notation, the
lepton doublets as $L_\alpha = (\nu_{\alpha L} \: e_{\alpha L})$,
where $\alpha = e,\mu,\tau$, and the right-handed charged leptons as
$E_\alpha = e_{\alpha R}$, where $\alpha = e,\mu,\tau$. Suppose that the part
of the scalar sector, which transforms non-trivially under the SM gauge
group, consists of two Higgs doublets $H_1$ and $H_2$, where $H_1$
couples to the neutrinos and $H_2$ to the charged
leptons.\footnote[2]{This can easily be achieved by imposing a
discrete ${\mathbb Z}_2$ symmetry under which $H_2$ and $E_\alpha$
($\alpha = e,\mu,\tau$) are odd and $H_1$ and the rest of the SM
fields are even.} (For simplicity and without loss of generality, the
quark sector will be left out in our entire discussion.) Let us first
restrict our discussion to the generation of the charged lepton
masses. In order to obtain the structure of the charged lepton mass
matrix from an underlying symmetry principle, we will further extend
the scalar sector by SM singlet scalar fields $\phi_i$ ($i =
1,2,\ldots,8$) and $\theta$ and assign the fields gauged horizontal
U(1) charges $Q_1$, $Q_2$, and $Q_3$ as follows:
$$
\begin{tabular}{c|c}
 & $(Q_1,Q_2,Q_3)$\\
\hline
$L_e$,$E_e$ & $(1,0,0)$\\
$L_\mu$,$L_\tau$,$E_\mu$,$E_\tau$ & $(0,1,0)$\\
\hline
$\phi_1$,$\phi_2$ & $(-1,1,2)$\\
$\phi_3$,$\phi_4$ & $(1,-1,2)$\\
$\phi_5$,$\phi_6$ & $(0,0,0)$\\
$\phi_7$,$\phi_8$ & $(0,0,1)$\\
$\theta$          & $(0,0,-1)$
\end{tabular}
$$
In the rest of the paper, it is always understood that the Higgs doublets $H_1$
and $H_2$ are total singlets under transformations of the additional
symmetries. Note that our model is kept anomaly-free, since the
fermions transform as vector-like pairs under the extra U(1) charges.
Next, the charges $(Q_1,Q_2)$ of the charged lepton-antilepton pairs
are given by:
$$
\begin{tabular}{c|c|ccc}
 & & $E_e$ & $E_\mu$ & $E_\tau$\\
\hline
 & $(Q_1,Q_2)$ & $(1,0)$ & $(0,1)$ & $(0,1)$\\
\hline
 & & & & \\[-3.5mm]
$\overline{L_e}$ & $(-1,0)$ & $(0,0)$ & $(-1,1)$ & $(-1,1)$\\
$\overline{L_\mu}$ & $(0,-1)$ & $(1,-1)$ & $(0,0)$ & $(0,0)$\\
$\overline{L_\tau}$ & $(0,-1)$ & $(1,-1)$ & $(0,0)$ & $(0,0)$\\
\end{tabular}
$$
We observe that these charges forbid dimension-four Yukawa coupling
terms for the coupling of the first generation to the second and third
generations. A realistic charged lepton mass matrix will arise if we,
in addition to the U(1) charges, introduce a set of discrete
symmetries ${\cal D}_i$ ($i = 1,2,\ldots,5$), which are, at this
level, not plagued with chiral anomalies.
The ${\mathbb Z}_4$ symmetry
\begin{eqnarray}
{\cal D}_1 & : & \left\{
\begin{matrix}
E_{\mu} \rightarrow {\rm i} E_{\mu}, & E_{\tau} \rightarrow {\rm i} E_{\tau},\\
\phi_3 \rightarrow - {\rm i} \phi_3, & \phi_4 \rightarrow - {\rm i}
\phi_4,\\ \phi_5 \rightarrow - {\rm i} \phi_5, & \phi_6 \rightarrow -
{\rm i} \phi_6,\\
\phi_7\rightarrow -{\rm i}\phi_7, & \phi_8\rightarrow -{\rm i} \phi_8
\end{matrix}\right.
\end{eqnarray}
forbids dimension-four Yukawa coupling terms in the
$\mu$-$\tau$-subsector of the charged lepton sector.
The ${\mathbb Z}_2$ symmetry
\begin{eqnarray}
{\cal D}_2 & : & \left\{
\begin{matrix}
E_e \rightarrow - E_e,\\
\phi_1 \rightarrow - \phi_1, & \phi_2 \rightarrow - \phi_2
\end{matrix}\right.
\end{eqnarray}
sets to leading order the $e$-$e$-element of the charged lepton mass
matrix equal to zero.
It has been pointed out that bimaximal leptonic mixing corresponds to
a permutation symmetry of the second and third generation \cite{moha001}.
Thus, we will introduce the three permutation symmetries
\begin{subequations}
\begin{eqnarray}
{\cal D}_3 & : & \left\{
\begin{matrix}
L_\mu \rightarrow - L_\mu, & E_\mu \rightarrow - E_\mu,\\
\phi_1 \leftrightarrow \phi_2, & \phi_3 \leftrightarrow \phi_4,
\end{matrix}\right. \\
{\cal D}_4 & : & \left\{
\begin{matrix}
L_\mu \rightarrow - L_\mu,\\
\phi_1 \leftrightarrow \phi_2, & \phi_5 \leftrightarrow \phi_6,\\
\phi_7 \rightarrow -\phi_7, &
\end{matrix}\right. \\
{\cal D}_5 & : & \left\{
\begin{matrix}
L_\mu \leftrightarrow L_\tau, & E_\mu \leftrightarrow E_\tau,\\
\phi_2 \rightarrow - \phi_2, & \phi_4 \rightarrow - \phi_4,\\
\phi_6 \rightarrow - \phi_6, & \phi_7\leftrightarrow\phi_8.
\end{matrix}\right.
\end{eqnarray}
\end{subequations}
Then, the most general charged lepton mass terms, which are invariant
under all symmetry transformations of our model, are given by the
higher-dimensional operators
\begin{equation}
{\cal L} = \overline{L_\alpha} H_2
\left[(Y_{\text{eff}}^1)_{\alpha\beta} +
(Y_{\text{eff}}^2)_{\alpha\beta}\right] E_\beta + \text{h.c.},
\end{equation}
where the relevant effective Yukawa interaction matrices $Y_{\text{eff}}^1$ and
$Y_{\text{eff}}^2$ are on the forms
\begin{subequations}\label{eq:Yeff}
\begin{eqnarray}
Y_{\text{eff}}^1 &=& {\small
\left( \begin{matrix} 0 & B(\phi_3 - \phi_4) & B (\phi_3 +
\phi_4)\\ A(\phi_1 - \phi_2) & C (\phi_5 - \phi_6) & 0\\ A
(\phi_1 + \phi_2) & 0 & C (\phi_5 + \phi_6) \end{matrix} \right)},
\nonumber\\\\
Y_{\text{eff}}^2 &=& \text{diag}(0, D \phi_7, D \phi_8).
\end{eqnarray}
\end{subequations}
Here, the dimensionful coefficients $A$, $B$, $C$, and $D$ are given by
\begin{subequations}\label{eq:ABCD}
\begin{eqnarray}
A &=& Y_a \frac{\theta^2}{M_1^3}, \\
B &=& Y_b \frac{\theta^2}{M_1^3}, \\
C &=& Y_c \frac{1}{M_1}, \\
D &=& Y_d \frac{\theta}{M_1^2},
\end{eqnarray}
\end{subequations}
where the quantities $Y_a$, $Y_b$, $Y_c$, and $Y_d$ are arbitrary order
unity coefficients and $M_1$ is the high mass scale of the intermediate
Froggatt--Nielsen states. Actually, in
\Sec~\ref{sec:chargedleptonmasses}, the mass scale $M_1$ will be
related to the breakdown scale of the extra symmetries by a small
expansion parameter.

\subsection{The scalar potential}
\label{sec:scalarpotential}

The most general renormalizable scalar potential, involving only the fields
$\phi_i$ $(i = 1,2,\ldots,6)$, which is invariant under
transformations of the horizontal symmetries given in
\Sec~\ref{sec:h_symm}, reads
\begin{eqnarray}
V &=& \mu_1^2 \left( |\phi_1|^2 + |\phi_2|^2 \right) \nonumber\\
&+& \mu_2^2 \left( |\phi_3|^2 + |\phi_4|^2 \right) \nonumber\\
&+& \mu_3^2 \left( |\phi_5|^2 + |\phi_6|^2 \right) \nonumber\\
&+& \kappa_1^2 \left( |\phi_1^\dagger \phi_1|^2 + |\phi_2^\dagger
\phi_2|^2 \right) \nonumber\\
&+& \kappa_2^2 \left( |\phi_3^\dagger \phi_3|^2 + |\phi_4^\dagger
\phi_4|^2 \right) \nonumber\\
&+& \kappa_3^2 \left( |\phi_5^\dagger \phi_5|^2 + |\phi_6^\dagger
\phi_6|^2 \right) \nonumber\\
&+& a |\phi_1^\dagger \phi_2|^2 + b |\phi_3^\dagger \phi_4|^2 + c
|\phi_5^\dagger \phi_6|^2 \nonumber\\
&+& d \left( |\phi_1^\dagger \phi_3|^2 + |\phi_2^\dagger \phi_3|^2 +
|\phi_1^\dagger \phi_4|^2 + |\phi_2^\dagger \phi_4|^2 \right)
\nonumber\\
&+& e \left( |\phi_1^\dagger \phi_5|^2 + |\phi_2^\dagger \phi_5|^2 +
|\phi_1^\dagger \phi_6|^2 + |\phi_2^\dagger \phi_6|^2 \right) \nonumber\\
&+& f \left( |\phi_3^\dagger \phi_5|^2 + |\phi_4^\dagger \phi_5|^2 +
|\phi_3^\dagger \phi_6|^2 + |\phi_4^\dagger \phi_6|^2 \right)
\nonumber
\end{eqnarray}
\begin{eqnarray}
&+& \lambda_1 \left[ \left(\phi_1^\dagger \phi_2\right)^2 +
\left(\phi_2^\dagger \phi_1\right)^2 \right] \nonumber\\
&+& \lambda_2 \left[ \left(\phi_3^\dagger \phi_4\right)^2 +
\left(\phi_4^\dagger \phi_3\right)^2 \right] \nonumber\\
&+& \lambda_3 \left[ \left(\phi_5^\dagger \phi_6\right)^2 +
\left(\phi_6^\dagger \phi_5\right)^2 \right] \nonumber\\
&+& m_1 \left(\phi_1^\dagger \phi_2 + \phi_2^\dagger \phi_1\right)
\left(\phi_3^\dagger \phi_4 + \phi_4^\dagger \phi_3\right) \nonumber\\
&+& m_2 \left(\phi_1^\dagger \phi_2 + \phi_2^\dagger \phi_1\right)
\left(\phi_5^\dagger \phi_6 + \phi_6^\dagger \phi_5\right) \nonumber\\
&+& m_3 \left(\phi_3^\dagger \phi_4 + \phi_4^\dagger \phi_3\right)
\left(\phi_5^\dagger \phi_6 + \phi_6^\dagger \phi_5\right), \nonumber\\
\end{eqnarray}
where all coefficients are real. Due to the symmetries of our model, the
remaining scalar fields enter relevant terms in the potential $V$ only
via quartic couplings in form of absolute squares of these fields, which means
that they can be combined into the coefficients $\mu_i$
$(i = 1,2,3)$. Therefore, we can choose the coefficients in the
potential to fulfill $\mu_i^2 < 0$ ($i = 1,2,3$), $\kappa_i > 0$ ($i =
1,2,3$), and $a,b,\ldots,f > 0$, which yield after spontaneous
symmetry breaking (SSB) non-vanishing vacuum expectation values (VEVs)
that satisfy
$$
|\langle \phi_1 \rangle| = |\langle \phi_2 \rangle|, \quad
|\langle \phi_3 \rangle| = |\langle \phi_4 \rangle|, \quad
|\langle \phi_5 \rangle| = |\langle \phi_6 \rangle|.
$$
If $\lambda_i < 0$ ($i = 1,2,3$), then we obtain pairwise relatively
real VEVs, \ie,
$$
\frac{\langle \phi_1 \rangle}{\langle \phi_2 \rangle},
\frac{\langle \phi_3 \rangle}{\langle \phi_4 \rangle}, \frac{\langle
\phi_5 \rangle}{\langle \phi_6 \rangle} \in \{-1,1\}. 
$$
Next, choosing $m_1 < 0$ and $m_2,m_3 > 0$, we obtain
$$
\frac{\langle \phi_1 \rangle}{\langle \phi_2 \rangle} \frac{\langle
\phi_3 \rangle}{\langle \phi_4 \rangle} = 1 \quad \mbox{and} \quad
\frac{\langle \phi_1 \rangle}{\langle \phi_2 \rangle} \frac{\langle
\phi_5 \rangle}{\langle \phi_6 \rangle} = - 1,
$$
\ie, the relative sign between $\langle \phi_1 \rangle$ and $\langle
\phi_2 \rangle$ is equal to the relative sign between $\langle \phi_3
\rangle$ and $\langle \phi_4 \rangle$ and opposite to the relative
sign between $\langle \phi_5 \rangle$ and $\langle \phi_6 \rangle$. In
\Sec~\ref{sec:chargedleptonmasses}, we will show that this alignment
mechanism reconciles the permutation symmetry ${\cal{D}}_5$ with an
approximate diagonal form of the charged lepton mass matrix.

\subsection{The charged lepton mass matrix}
\label{sec:chargedleptonmasses}

Suppose that the SM singlet scalar fields aquire their VEVs at a high
mass scale and thereby give rise to a small expansion parameter
\begin{equation}\label{eq:expansionparameter}
\epsilon \simeq \frac{\langle \phi_i \rangle}{M_1}
\simeq \frac{\langle \theta \rangle}{M_1} \simeq 10^{-1},
\end{equation}
where $i = 1,2,\ldots,8$. Such small hierarchies can arise from large
hierarchies in supersymmetric theories when the scalar fields aquire
their VEVs along a ``D-flat'' direction
\cite{Witten:1981kv}. As a consequence of the
permutation symmetries ${\cal D}_3$, ${\cal D}_4$, and ${\cal D}_5$, 
the lowest energy state is two-fold degenerate. Applying the results
of \Sec~\ref{sec:scalarpotential} and inserting
\eq~(\ref{eq:expansionparameter}) into \eqs~(\ref{eq:ABCD}) and the result
thereof into \eqs~(\ref{eq:Yeff}), we obtain
the two possible charged lepton mass matrices
\begin{equation}
M_\ell \simeq m_\tau
\left( \begin{matrix} 0 & \epsilon^2 & 0\\ \epsilon^2 &
\epsilon & 0\\ 0 & 0 & 1 \end{matrix} \right)
\label{eq:Ml}
\end{equation}
and
\begin{equation}
M_\ell \simeq m_\tau \left( \begin{matrix} 0 & 0 & \epsilon^2\\ 0 & 1 & 0\\
\epsilon^2 & 0 & \epsilon \end{matrix} \right),
\end{equation}
where $m_\tau$ is the tau mass and only the order of magnitude of the
matrix elements have been indicated.
Note that a permutation of the second and third generations, $L_\mu
\leftrightarrow L_\tau$, $E_\mu \leftrightarrow E_\tau$, leads from
one solution to the other. Let us take the first one. Diagonalization
of $M_\ell$ [in \eq~(\ref{eq:Ml})] then gives for the charged lepton
masses the order-of-magnitude relations\footnote[3]{The charged lepton
mass spectrum is: $m_e \equiv |\lambda_1|$, $m_\mu \equiv
|\lambda_2|$, $m_\tau \equiv |\lambda_3|$, where $\lambda_1 =
\tfrac{\epsilon}{2} \left( 1 - \sqrt{1 + 4 \epsilon^2} \right) m_\tau
\simeq - \epsilon^3 m_\tau$, $\lambda_2 = \tfrac{\epsilon}{2} \left( 1
+ \sqrt{1 + 4 \epsilon^2} \right) m_\tau \simeq \epsilon m_\tau$, and
$\lambda_3 = m_\tau$ are the eigenvalues of the matrix $M_\ell$.}
$$
m_e/m_\mu \simeq \epsilon^2 \simeq 10^{-2} \quad \mbox{and} \quad
m_\mu/m_\tau \simeq \epsilon \simeq 10^{-1},
$$
which approximately fit the experimentally observed values
[\ie, $(m_e/m_\mu)_{\rm exp} \simeq 4.8 \cdot 10^{-3}$ and
$(m_\mu/m_\tau)_{\rm exp} \simeq 5.9 \cdot 10^{-2}$ \cite{groo00}]. Here $m_e$
and $m_\mu$ are the electron and muon masses, respectively. The
charged lepton mass matrix $M_\ell$ is diagonalized by a rotation of the
left-handed charged lepton fields in the 1-2-plane by an angle
$\theta^\ell_{12}\simeq 6^\circ$ ($\theta^\ell_{12} = \epsilon -
\tfrac{4}{3} \epsilon^3 + {\cal O}(\epsilon^5)$, where $\epsilon
\simeq 0.1$, $\theta^\ell_{13} = 0$, and $\theta^\ell_{23} = 0$),
which will finally give a contribution to all leptonic mixing angles
(see \App~\ref{app:U}).

\section{The neutrino mass matrix}
\label{sec:neutrinomassmatrix}

Let us now turn our discussion to the neutrino mass matrix. As intermediate
Froggatt--Nielsen states we will assume two heavy SM singlet Dirac
fermions $F_1$ and $F_2$, which have masses of the order $M_1$. In
order to account for the smallness of the neutrino masses, we will
furthermore introduce three SM singlet Dirac fermions $N_e$, $N_\mu$,
and $N_\tau$, which have masses of the order of some relevant high
mass scale $M_2$.
{}From the assignment of the charges $(Q_1,Q_2)$ to the
lepton doublets, the structure of the $(Q_1,Q_2)$ charges associated
with the effective neutrino mass matrix follows immediately:
$$
\begin{tabular}{c|c|ccc}
 & & $L_e$ & $L_\mu$ & $L_\tau$\\
\hline
 & $(Q_1,Q_2)$ & $(1,0)$ & $(0,1)$ & $(0,1)$\\
\hline
 & & & & \\[-3.5mm]
$\overline{L_e^c}$ & $(1,0)$ & $(2,0)$ & $(1,1)$ & $(1,1)$\\
$\overline{L_\mu^c}$ & $(0,1)$ & $(1,1)$ & $(0,2)$ & $(0,2)$\\
$\overline{L_\tau^c}$ & $(0,1)$ & $(1,1)$ & $(0,2)$ & $(0,2)$\\
\end{tabular}
$$
Note that the given representation content so far forbids any neutrino
mass term. We therefore introduce the additional SM singlet scalar
fields $\phi_9$, $\phi_{10}$, $\phi_{11}$, and $\phi_{12}$ and we assign the
charges $Q_1$, $Q_2$, and $Q_3$ to the fields as follows:
$$
\begin{tabular}{c|c}
 & $(Q_1,Q_2,Q_3)$\\
\hline
$N_e$ & $(1,0,0)$\\
$N_\mu,N_\tau$ & $(0,1,0)$\\
$F_1$ & $(1,0,0)$\\
$F_2$ & $(-1,0,1)$\\
\hline
$\phi_9$,$\phi_{10}$ & $(-1,-1,0)$\\
$\phi_{11}$ & $(-2,0,1)$\\
$\phi_{12}$ & $(0,0,0)$\\
\end{tabular}
$$
Note that our model is again kept free from chiral
anomalies, since the heavy fermions are vector representations under
transformations of all U(1) charges. The Dirac neutrino fields and the
scalar fields
transform under the discrete symmetries ${\cal D}_3$, ${\cal D}_4$,
${\cal D}_5$, and the additional discrete symmetry ${\cal D}_6$ as
\begin{subequations}
\begin{eqnarray}
{\cal D}_3&:& \ldots, \; N_\mu \rightarrow - N_\mu,\;
 \phi_9 \rightarrow - \phi_9, \\
{\cal D}_4&:& \ldots, \; N_\mu \rightarrow - N_\mu,\;
 \phi_9 \rightarrow - \phi_9, \\
{\cal D}_5&:& \ldots, \; N_\mu \leftrightarrow N_\tau,\;
 \phi_9 \leftrightarrow \phi_{10}, \\
{\cal D}_6&:& \left\{
\begin{matrix}
N_e \rightarrow {\rm i} N_e,\\
\phi_{11} \rightarrow - {\rm i} \phi_{11}, & \phi_{12} \rightarrow
{\rm i} \phi_{12}.
\end{matrix}\right.
\end{eqnarray}
\end{subequations}
It is easily verified that the results
for the charged lepton mass matrix remain unchanged by this new representation
content.
The leading order tree-level realizations of the higher-dimensional
operators, which generate the neutrino masses, are shown in
\figs~\ref{fig:dimsix} and \ref{fig:dimeight}.
\begin{figure}
\begin{center}
\begin{fmffile}{dimsix}
\begin{fmfgraph*}(150,20)\fmfpen{thin}
 \fmfstraight
 \fmfleft{i1,i2}\fmfright{o1,o2}
 \fmf{phantom}{i1,v1,v2,v3,v4,v5,o1}
 \fmf{plain,label.side=right,label=\small{$\overline{L^c_e}$}}{v6,i2}
 \fmf{plain,label.side=left,label=\small{$F_1^c$}}{v6,v7}
 \fmf{plain,label.side=left,label=\small{$\overline{F_1^c}$}}{v7,v8}
 \fmf{plain,label.side=left,label=\small{$N_\alpha$}}{v8,v9}
 \fmf{plain,label.side=left,label=\small{$\overline{N_\alpha}$}}{v9,v10}
 \fmf{plain,label.side=left,label=\small{$L_\alpha$}}{v10,o2}
 \fmf{dashes,tension=0}{v1,v6}
 \fmf{dashes,tension=0}{v3,v8}
 \fmf{dashes,tension=0}{v5,v10}
 \fmfv{decor.shape=circle,decor.filled=full,decor.size=2thin}{v6,v8,v10}
 \fmfv{decor.shape=circle,decor.filled=full,decor.size=2thin,
       label=\small{$H_1$}}{v1}
 \fmfv{decor.shape=circle,decor.filled=full,decor.size=2thin,
       label=\small{$\phi_\alpha$}}{v3}
 \fmfv{decor.shape=circle,decor.filled=full,decor.size=2thin,
       label=\small{$H_1$}}{v5}
 \fmfv{decor.shape=cross,decor.size=10}{v7,v9}
\end{fmfgraph*}
\end{fmffile}
\caption{\small{The dimension six operator for $\alpha = \mu,\tau$ and
$\phi_\mu \equiv \phi_9$, $\phi_\tau \equiv \phi_{10}$, generating the
$e$-$\mu$- and $e$-$\tau$-elements in the effective neutrino mass matrix.}}
\label{fig:dimsix}
\end{center}
\end{figure}
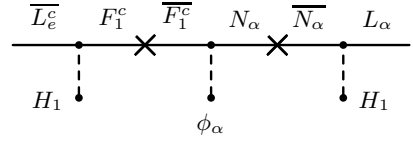
\begin{figure}
\begin{center}
\begin{fmffile}{dimeight1}
\begin{fmfgraph*}(210,20)\fmfpen{thin}
 \fmfstraight
 \fmfleft{i1,i2}\fmfright{o1,o2}
 \fmf{phantom}{i1,v1,v2,v3,v4,v5,v6,v7,v8,v9,o1}
 \fmf{plain,label.side=right,label=\small{$\overline{L^c_e}$}}{v10,i2}
 \fmf{plain,label.side=left,label=\small{$F_1^c$}}{v10,v11}
 \fmf{plain,label.side=left,label=\small{$\overline{F_1^c}$}}{v11,v12}
 \fmf{plain,label.side=left,label=\small{$F_2$}}{v12,v13}
 \fmf{plain,label.side=left,label=\small{$\overline{F_2}$}}{v13,v14}
 \fmf{plain,label.side=left,label=\small{$N_e$}}{v14,v15}
 \fmf{plain,label.side=left,label=\small{$\overline{N_e}$}}{v15,v16}
 \fmf{plain,label.side=left,label=\small{$F_1$}}{v16,v17}
 \fmf{plain,label.side=left,label=\small{$\overline{F_1}$}}{v17,v18}
 \fmf{plain,label.side=left,label=\small{$L_e$}}{v18,o2}
 \fmf{dashes,tension=0}{v1,v10}
 \fmf{dashes,tension=0}{v3,v12}
 \fmf{dashes,tension=0}{v5,v14}
 \fmf{dashes,tension=0}{v7,v16}
 \fmf{dashes,tension=0}{v9,v18}
 \fmfv{decor.shape=circle,decor.filled=full,decor.size=2thin}{v10,v12,v14,v16,v18}
 \fmfv{decor.shape=circle,decor.filled=full,decor.size=2thin,
       label=\small{$H_1$}}{v1}
 \fmfv{decor.shape=circle,decor.filled=full,decor.size=2thin,
       label=\small{$\theta$}}{v3}
 \fmfv{decor.shape=circle,decor.filled=full,decor.size=2thin,
       label=\small{$\phi_{11}$}}{v5}
 \fmfv{decor.shape=circle,decor.filled=full,decor.size=2thin,
       label=\small{$\phi_{12}$}}{v7}
 \fmfv{decor.shape=circle,decor.filled=full,decor.size=2thin,
       label=\small{$H_1$}}{v9}
 \fmfv{decor.shape=cross,decor.size=8}{v11,v13,v15,v17}
\end{fmfgraph*}
\end{fmffile}
\end{center}
${}$\\
\begin{center}
\begin{fmffile}{dimeight2}
\begin{fmfgraph*}(210,20)\fmfpen{thin}
 \fmfstraight
 \fmfleft{i1,i2}\fmfright{o1,o2}
 \fmf{phantom}{i1,v1,v2,v3,v4,v5,v6,v7,v8,v9,o1}
 \fmf{plain,label.side=right,label=\small{$\overline{L^c_e}$}}{v10,i2}
 \fmf{plain,label.side=left,label=\small{$F_1^c$}}{v10,v11}
 \fmf{plain,label.side=left,label=\small{$\overline{F_1^c}$}}{v11,v12}
 \fmf{plain,label.side=left,label=\small{$N_e^c$}}{v12,v13}
 \fmf{plain,label.side=left,label=\small{$\overline{N_e^c}$}}{v13,v14}
 \fmf{plain,label.side=left,label=\small{$F_2^c$}}{v14,v15}
 \fmf{plain,label.side=left,label=\small{$\overline{F_2^c}$}}{v15,v16}
 \fmf{plain,label.side=left,label=\small{$F_1$}}{v16,v17}
 \fmf{plain,label.side=left,label=\small{$\overline{F_1}$}}{v17,v18}
 \fmf{plain,label.side=left,label=\small{$L_e$}}{v18,o2}
 \fmf{dashes,tension=0}{v1,v10}
 \fmf{dashes,tension=0}{v3,v12}
 \fmf{dashes,tension=0}{v5,v14}
 \fmf{dashes,tension=0}{v7,v16}
 \fmf{dashes,tension=0}{v9,v18}
 \fmfv{decor.shape=circle,decor.filled=full,decor.size=2thin}{v10,v12,v14,v16,v18}
 \fmfv{decor.shape=circle,decor.filled=full,decor.size=2thin,
       label=\small{$H_1$}}{v1}
 \fmfv{decor.shape=circle,decor.filled=full,decor.size=2thin,
       label=\small{$\phi_{12}$}}{v3}
 \fmfv{decor.shape=circle,decor.filled=full,decor.size=2thin,
       label=\small{$\phi_{11}$}}{v5}
 \fmfv{decor.shape=circle,decor.filled=full,decor.size=2thin,
       label=\small{$\theta$}}{v7}
 \fmfv{decor.shape=circle,decor.filled=full,decor.size=2thin,
       label=\small{$H_1$}}{v9}
 \fmfv{decor.shape=cross,decor.size=8}{v11,v13,v15,v17}
\end{fmfgraph*}
\end{fmffile}
\caption{\small{The dimension eight operators, generating the
$e$-$e$-element in the effective neutrino mass
matrix.}}\label{fig:dimeight}
\end{center}
\end{figure}
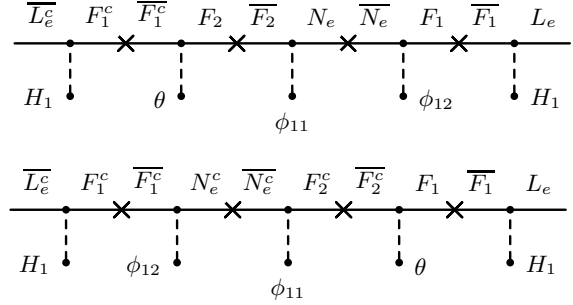
After SSB, the effective neutrino mass
matrix $M_\nu$ will be on the approximate bimaximal mixing form \cite{barg98}
\begin{equation}\label{eq:neutrinomassmatrix}
M_\nu = \left( \begin{matrix} A' & B' & -B'\\ B' & 0 & 0\\ -B' & 0 & 0
\end{matrix} \right),
\end{equation}
where
\begin{subequations}\label{eq:entries}
\begin{eqnarray}
A' &=& Y'_a \frac{\langle H_1 \rangle^2}{M_2} \frac{\langle \phi_{11}
\rangle \langle \phi_{12} \rangle \langle \theta \rangle}{(M_1)^3},
\label{eq:entryA}\\
B' &=& Y'_b \frac{\langle H_1 \rangle^2}{M_2} \frac{\langle \phi_9
\rangle}{M_1}, \label{eq:entryB}
\end{eqnarray}
\end{subequations}
and the quantities $Y'_a$ and $Y'_b$ are arbitrary order unity
coefficients. Note that the permutation symmetry ${\cal D}_5$
establishes $|\langle \phi_9 \rangle| = |\langle \phi_{10} \rangle|$,
and hence, the magnitudes of the $e$-$\mu$- and $e$-$\tau$-entries in the
effective neutrino mass matrix $M_\nu$ are exactly degenerate. The
possible relative phase $\varphi$ between these entries can be
eliminated by, \eg, the field redefinition $L_\tau \rightarrow {\rm
e}^{{\rm i} \varphi} L_\tau$, since it does not affect the mixing
angles in the charged lepton sector. If we assume that all SM singlet
scalar fields aquire their VEVs at the breakdown scale of the
additional horizontal symmetries, then \eq~(\ref{eq:expansionparameter}) 
also applies to the indices $i=9,10,11,12$. Therefore, we can
parameterize the effective neutrino mass matrix $M_\nu$ with the same
expansion parameter $\epsilon$ that was used for the charged lepton
mass matrix $M_\ell$ and we find the relative suppression ratio $A'/B' \simeq
\epsilon^2\simeq 10^{-2}$ from \eq~(\ref{eq:entries}). Thus,
diagonalizing the neutrino mass matrix $M_\nu$ in
\eq~(\ref{eq:neutrinomassmatrix}), we obtain the following spectrum of
the neutrino masses\footnote[4]{The neutrino mass spectrum is: $m_1 \equiv
|\lambda'_1|$, $m_2 \equiv |\lambda'_2|$, $m_3 \equiv |\lambda'_3|$,
where $\lambda'_1 \simeq - B' \sqrt{2}$, $\lambda'_2 \simeq B'
\sqrt{2}$, and $\lambda'_3 = 0$ are the eigenvalues of the matrix $M_\nu$.}
$$
m_1 \simeq m_2 \quad \mbox{and} \quad m_3 = 0,
$$
which is on the inverted hierarchical form (\ie, $m_3 \ll m_1 \simeq m_2
\quad \Rightarrow \quad 0 \simeq |\Delta m_{21}^2| \ll |\Delta
m_{32}^2| \simeq |\Delta m_{31}^2|$).
Using the MSW LMA and atmospheric neutrino mass squared
differences\footnote[5]{Here $\Delta m_{ab}^2 = {\lambda'}_a^2 -
{\lambda'}_b^2$, \ie, $|\Delta m_{21}^2| \simeq 2 \sqrt{2} |A'B'|$ and
$|\Delta m_{32}^2| \simeq |\Delta m_{31}^2| \simeq 2 |{B'}^2|$.}
\cite{Bahcall:2001cb,Toshito:2001dk}
\begin{eqnarray}
\Delta m_\odot^2 & \equiv & |\Delta m_{21}^2| \simeq 3.7 \cdot 10^{-5}
\, {\rm eV}^2 \sim 10^{-5} \, {\rm eV}^2,
\nonumber\\ 
\Delta m_{\text{atm}}^2 & \equiv & |\Delta m^2_{32}| \simeq 2.5 \cdot
10^{-3} \, {\rm eV}^2 \sim 10^{-3} \, {\rm eV}^2,
\nonumber
\end{eqnarray}
and a VEV of the order of the electroweak scale $\langle H_1 \rangle
\simeq 10^2 \, {\rm GeV}$, we obtain from \eqs~(\ref{eq:entries}) the
high mass scale $M_2 \simeq 10^{14} \, {\rm GeV}$ (as well as $m_1 \simeq m_2
\simeq 0.05 \, {\rm eV}$ and $\epsilon \simeq 0.1$).
The entry $A'$ of the matrix in \eq~(\ref{eq:neutrinomassmatrix}) induces a
deviation from maximal solar mixing, which is of the order
$0.1^\circ$ ($\theta^\nu_{12} = \tfrac{\pi}{4} + 
\tfrac{1}{4 \sqrt{2}} \epsilon^2 - \tfrac{1}{96 \sqrt{2}} \epsilon^4 +
{\cal O}(\epsilon^{10})$, where $\epsilon \simeq \sqrt{A'/B'} \simeq
0.1$, $\theta^\nu_{13} = 0$, and $\theta^\nu_{23} = 45^\circ$). This
deviation can be neglected in comparison with the contribution coming
from the charged lepton sector, which is about $6^\circ$ (see
\Sec~\ref{sec:chargedleptonmasses}), resulting in
a change of the mixing angles $\theta_{12}$ and $\theta_{13}$ by
approximately $4^\circ$, while the atmospheric mixing angle
$\theta_{23}$ practically stays maximal (see \App~\ref{app:U}). Thus,
the leptonic mixing angles are
$$
\theta_{12} \simeq 41^\circ, \quad \theta_{13} \simeq 4^\circ, \quad \mbox{and}
\quad \theta_{23} \simeq 45^\circ.
$$
Hence, our model predicts
the charged lepton mass hierarchy spectrum, an inverted neutrino mass
hierarchy spectrum, bilarge leptonic mixing, as well as it reproduces
the mass squared differences to lie within the ranges preferred by the MSW LMA
solution\footnote[6]{@ 99.73\% C.L.: $1.9 \cdot
10^{-5} \, {\rm eV}^2 \lesssim \Delta m_\odot^2 \lesssim 2.7 \cdot
10^{-4} \, {\rm eV}^2$ \cite{Gonzalez-Garcia:2002dz}; best-fit:
$\Delta m_\odot^2 \simeq 3.7 \cdot 10^{-5} \, {\rm eV}^2$
\cite{Bahcall:2001cb}} and atmospheric neutrino data\footnote[7]{@ 90\% C.L.:
$1.6 \cdot 10^{-3} \, {\rm eV}^2 \lesssim \Delta m_{\rm atm}^2
\lesssim 4.0 \cdot 10^{-3} \, {\rm eV}^2$; best-fit: $\Delta m_{\rm
atm}^2 \simeq 2.5 \cdot 10^{-3} \, {\rm eV}^2$
\cite{Toshito:2001dk}}. In particular, it yields a significant
deviation from maximal solar mixing. However, the solar mixing angle
is bounded from below by approximately $41^\circ$ and it is therefore
still too close to maximal to be in the 95\% (or 99.73\%) confidence
level region of the MSW LMA solution\footnote[8]{@ 99.73\% C.L.: $0.22 \lesssim
\tan^2 \theta_{12} \lesssim 0.71 \quad \Rightarrow \quad 25^\circ
\lesssim |\theta_{12}| \lesssim 40^\circ$
\cite{Gonzalez-Garcia:2002dz}; best-fit: $\tan^2 \theta_{12} 
\simeq 0.37 \quad \Rightarrow \quad |\theta_{12}| \simeq 31^\circ$
\cite{Bahcall:2001cb}} \cite{bahc01,Bahcall:2001cb}.

\section{Summary and conclusions}
\label{sec:S&C}

In summary, we have presented a model built upon extra fields, horizontal
(flavor) symmetries, and higher-dimensional operators including the
Froggatt--Nielsen mechanism. This model naturally yields the well-known mass
matrix textures
$$
\left( \begin{matrix} 0 & \epsilon^2 & 0\\ \epsilon^2 & \epsilon
& 0\\ 0 & 0 & 1 \end{matrix} \right) \quad \mbox{and} \quad \left(
\begin{matrix} \epsilon^2 & 1 & -1\\ 1 & 0 & 0\\ -1 & 0 & 0
\end{matrix} \right)
$$
for the charged leptons and the neutrinos, respectively, which involve the
same small expansion parameter $\epsilon \simeq 0.1$. These textures reproduce
the mass hierarchy among the charged leptons very accurately as well as they
give rise to the most probable values of the mass squared differences
for the neutrinos coming from solar and atmospheric neutrino data. In
addition, assuming no ${\cal CP}$ violation, \ie, $\delta = 0$, the
model gives bilarge leptonic mixing, \ie, $\theta_{12} \simeq
41^\circ$, $\theta_{13} \simeq 4^\circ$, and $\theta_{23} \simeq
45^\circ$, which is in very good agreement with the present
experimental data. The mixing angle $\theta_{12}$ is on the borderline
of being compatible with the MSW LMA solution, which has, however, been further
strengthened by recent SNO results \cite{ahma02} and which will,
nevertheless, soon be even better tested by the KamLAND experiment
\cite{Piepke:2001tg},
whereas $\theta_{13}$ is well below the CHOOZ upper bound, and $\theta_{23}$
fits perfectly the best-fit value from the Super-Kamiokande
collaboration of their atmospheric neutrino data.

As a final conclusion, we have shown in \App~\ref{app:U} that the
leptonic mixing angles are all dependent on the ``solar'' mixing angle
$\theta^\ell_{12}$ in the charged lepton sector in a non-trivial way.

\section*{Acknowledgments}

We would like to thank Martin Freund, Patrick Huber, and Manfred
Lindner for useful discussions.

This work was supported by the Swedish Foundation for International
Cooperation in Research and Higher Education (STINT) [T.O.], the 
Wenner-Gren Foundations [T.O.], the Magnus Bergvall Foundation
(Magn. Bergvalls Stiftelse) [T.O.], and the ``Sonderforschungsbereich
375 f{\"u}r Astro-Teilchenphysik der Deutschen
Forschungsgemeinschaft'' [T.O. and G.S.].

\begin{appendix}

\section{The leptonic mixing matrix - A special case}
\label{app:U}

The leptonic mixing matrix can be written as
\begin{eqnarray}
U &=& O_{23}(\theta_{23}) U_{13}(\theta_{13},\delta)
O_{12}(\theta_{12}) \nonumber\\
&=& \left(\begin{matrix} 1 & 0 & 0\\ 0 & C_{23} & S_{23}\\ 0 & - S_{23} &
C_{23} \end{matrix}\right) \nonumber\\
&\times& \left(\begin{matrix} C_{13} & 0 & S_{13}
{\rm e}^{- {\rm i} \delta}\\ 0 & 1 & 0\\ - S_{13} {\rm e}^{{\rm i}
\delta} & 0 & C_{13} \end{matrix}\right) \nonumber\\
&\times& \left(\begin{matrix} C_{12} & S_{12} & 0\\
- S_{12} & C_{12} & 0\\ 0 & 0 & 1 \end{matrix}\right),
\label{eq:U}
\end{eqnarray}
where $C_{ab} \equiv \cos \theta_{ab}$, $S_{ab} \equiv \sin
\theta_{ab}$, and $O_{ab}(\theta_{ab})$ is a rotation by an angle
$\theta_{ab}$ in the $ab$-plane. If $\delta = 0$, then
$U_{13}(\theta_{13},0) = O_{13}(\theta_{13})$.
Assuming that all the ${\cal CP}$ phases are zero, \ie, $\delta =
\delta^\ell = \delta^\nu = 0$, then
the charged lepton and neutrino mixing matrices, $U^\ell$ and $U^\nu$,
can, of course, also be written in the same form as the above complete
leptonic mixing matrix, $U$, \ie, we have
\begin{subequations}
\begin{equation}
U^\ell = O^\ell_{23}(\theta^\ell_{23})
\underbrace{U^\ell_{13}(\theta^\ell_{13},
\delta^\ell)}_{O^\ell_{13}(\theta^\ell_{13})} O^\ell_{12}(\theta^\ell_{12}),
\end{equation}
\begin{equation}
U^\nu = O^\nu_{23}(\theta^\nu_{23}) \underbrace{U^\nu_{13}(\theta^\nu_{13},
\delta^\nu)}_{O^\nu_{13}(\theta^\nu_{13})} O^\nu_{12}(\theta^\nu_{12}).
\end{equation}
\label{eq:UellUnu}
\end{subequations} 
Thus, inserting \eqs~(\ref{eq:UellUnu}) into the
definition of the leptonic mixing matrix, $U = (U_{\alpha a} \equiv
{U^\ell}^\dagger U^\nu$, we find that
\begin{eqnarray}
U &=& {O^\ell_{12}(\theta^\ell_{12})}^T
{O^\ell_{13}(\theta^\ell_{13})}^T
{O^\ell_{23}(\theta^\ell_{23})}^T \nonumber\\
&\times& O^\nu_{23}(\theta^\nu_{23}) O^\nu_{13}(\theta^\nu_{13})
O^\nu_{12}(\theta^\nu_{12}).
\end{eqnarray}
Furthermore, assuming that we have only a small mixing coming from
the mixing angle $\theta^\ell_{12}$ in the charged lepton sector
($\theta^\ell_{13} = 0$, $\theta^\ell_{23} = 0$) and bimaximal mixing
in the neutrino sector ($\theta^\nu_{12} = 45^\circ$, $\theta^\nu_{13}
= 0$, $\theta^\nu_{23} = 45^\circ$), we then obtain
\begin{eqnarray}
U &=& O^\ell_{12}(\theta^\ell_{12})^T O^\nu_{23}(\theta^\nu_{23} =
45^\circ) O^\nu_{12}(\theta^\nu_{12} = 45^\circ) \nonumber\\
&=& {\tiny \left(\begin{matrix} C^\ell_{12}/\sqrt{2} + S^\ell_{12}/2 &
C^\ell_{12}/\sqrt{2} - S^\ell_{12}/2 & - S^\ell_{12}/\sqrt{2}\\
S^\ell_{12}/\sqrt{2} - C^\ell_{12}/2 & S^\ell_{12}/\sqrt{2} +
C^\ell_{12}/2 & C^\ell_{12}/\sqrt{2}\\ 1/2 & - 1/2 & 1/\sqrt{2}
\end{matrix}\right).
} \nonumber\\ \label{eq:UO12}
\end{eqnarray}
The mixing angles (in the standard parameterization) of a $3 \times 3$
orthogonal mixing matrix can be read off as follows \cite{Ohlsson:1999xb}:
\begin{eqnarray}
\theta_{12} &=& \arctan \frac{U_{e2}}{U_{e1}}, \label{eq:t12}\\
\theta_{13} &=& \arcsin U_{e3},\\
\theta_{23} &=& \arctan \frac{U_{\mu 3}}{U_{\tau 3}}. \label{eq:t23}
\end{eqnarray}
Thus, inserting the appropriate matrix elements of the matrix $U$ in
\eq~(\ref{eq:UO12}) into \eqs~(\ref{eq:t12}) - (\ref{eq:t23}), we finally
obtain
\begin{eqnarray}
\theta_{12} &=& \arctan \frac{\cos \theta^\ell_{12} -
\tfrac{1}{\sqrt{2}}\sin \theta^\ell_{12}}{\cos \theta^\ell_{12} +
\tfrac{1}{\sqrt{2}}\sin \theta^\ell_{12}},\\
\theta_{13} &=& - \arcsin \left( \frac{1}{\sqrt{2}} \sin
\theta^\ell_{12} \right),\\
\theta_{23} &=& \arctan \cos \theta^\ell_{12}.
\end{eqnarray}
When $\theta^\ell_{12}$ is small ($\theta^\ell_{12} \ll 1$), we
have\footnote[9]{Introducing a small deviation $\eta$ from maximal
solar mixing in the neutrino sector (\ie, $\theta^\nu_{12} =
45^\circ \quad \to \quad \theta^\nu_{12} = 45^\circ + \eta$, where
$\eta \ll 1$), we find that $\theta_{12} = \tfrac{\pi}{4} + \eta -
\tfrac{1}{\sqrt{2}} {\theta^\ell_{12}} - \tfrac{1}{6 \sqrt{2}}
{\theta^\ell_{12}}^3 + {\cal O}({\theta^\ell_{12}}^5)$, whereas
$\theta_{13}$ and $\theta_{23}$ remain unchanged.}
\begin{equation}
\theta_{12} = \frac{\pi}{4} - \frac{1}{\sqrt{2}} \theta^\ell_{12}
- \frac{1}{6 \sqrt{2}} {\theta^\ell_{12}}^3 + {\cal
O}({\theta^\ell_{12}}^5),
\end{equation}
\begin{eqnarray}
\theta_{13} &=& - \frac{1}{\sqrt{2}} \theta^\ell_{12} + \frac{1}{12
\sqrt{2}} {\theta^\ell_{12}}^3 + {\cal O}({\theta^\ell_{12}}^5), \nonumber\\\\
\theta_{23} &=& \frac{\pi}{4} - \frac{1}{4} {\theta^\ell_{12}}^2 -
\frac{1}{24} {\theta^\ell_{12}}^4 + {\cal O}({\theta^\ell_{12}}^6). \nonumber\\
\end{eqnarray}
Note that all leptonic mixing angles receive contribution from the
small mixing angle $\theta^\ell_{12}$ in the charged lepton
sector. Furthermore, we observe that $\theta_{12}$ has first order
corrections in $\theta^\ell_{12}$, whereas $\theta_{23}$ has only
second order corrections in $\theta^\ell_{12}$. The mixing angle
$\theta_{13}$ is directly proportional to the mixing angle
$\theta^\ell_{12}$ (when $\theta^\ell_{12}$ is small), which means
that if $\theta^\ell_{12}$ is small, then $\theta_{13}$ will also be
small. In fact, $|\theta^\ell_{12}| \lesssim 13.1^\circ$ has to be
fulfilled in order for the mixing angle $\theta_{13}$ to be below the
CHOOZ upper bound $\sin^2 2 \theta_{13} \lesssim 0.10$ (\ie, $|\theta_{13}|
\lesssim 9.2^\circ$) \cite{Apollonio:1998xe}.

\end{appendix}

\end{document}